\documentclass[final,5p,times,twocolumn,authoryear]{elsarticle}
\usepackage{amssymb}
\usepackage{lipsum}
\usepackage{amsmath}
\usepackage{balance}
\usepackage[dvipsnames]{xcolor}
\usepackage{natbib}
\setcitestyle{numbers,comma,square}
\usepackage{makecell}
\usepackage{url}
\usepackage[labelfont=bf,justification=raggedright,singlelinecheck=false]{caption}
\begin{document}

\begin{frontmatter}



\title{Aluminium nanoparticle-based ultra-wideband high-performance polarizer}

\author[1]{Md. Shariful Islam}
%\ead{0421062308@eee.buet.ac.bd}
%\cormark[]

\affiliation[1]{organization={Department of Electrical and Electronic Engineering, Bangladesh University of Engineering and Technology},%Department and Organization
            %addressline={}, 
            city={Dhaka},
            postcode={1205}, 
            %state={},
            country={Bangladesh}}

\author[1]{Ahmed Zubair}
%[orcid=0000-0002-1833-2244]
%\cormark[cor1]
\ead{ahmedzubair@eee.buet.ac.bd}

\cortext[cor1]{Corresponding author}

\begin{abstract}
The polarizer-based device industry is expanding quickly, requiring high-quality research on nanoscale wideband polarizers. Here, we investigated the possibility of utilizing Al dimer nanostructures on broad-band polarizers. Metals are always considered promising candidates for reflection-based polarizer development because of their high extinction ratio. This study proposes a nanoparticle polarizer comprised of semi-immersed Al nanodimers with a 200 nm radius on a CaF$_2$ substrate. Our proposed polarizer has effective polarization anisotropy in the near-infrared (NIR) and THz range. This study includes calculating performance parameters for the extraction of the proposed polarizer, including insertion loss, extinction ratio (ER), Mueller matrix values, and polarization ellipse diagram. The finite-difference time-domain (FDTD) simulation-based results suggested obtaining more than 55 dB extinction ratio for the 0.2 to 9 THz range. The average extinction ratio and insertion loss over the 1--1665 $\mu$m wavelength were 29.01 dB and $\sim$1 dB, respectively. We have reviewed recent reports of similar nanoparticle and wire grid-based polarizers to evaluate our Al nanodimer-based polarizer and performed a comparative analysis. The idea of Al dimer and the insight gained from the results extracted from the rigorous simulation report suggested a great opportunity for developing micro-scale metallic wideband polarizers.
\end{abstract}

%Graphical abstract
% \includegraphics[width=1\textwidth]{Graphical abstract.png}

% \begin{highlights}
% \item We proposed a novel polarizer design based on Al dimer nanostructures submerged in CaF2, which opened up an excellent avenue for developing a micro-scale metallic reflection-based wideband polarizer.

% \item Extinction ratio of $>$55 dB was obtained for the 0.2 to 9 THz range.

% \item Excellent average extinction ratio of 29.01 dB and low insertion loss of $\sim$ 1 dB were achieved over an ultrabroadband wavelength range of 1-1665 $\mu$m.

% \end{highlights}


\end{frontmatter}

%\tableofcontents



\section{Introduction}
\label{introduction}

The pioneering work of Malus expedited the development of the polarizer.  %\cite{malus_1809_malus_law}. 
In the early twentieth century, several observations and experiments, such as Herapath, Ambronn, and C.E. Hall's works\,\cite{H.Land_1951_Sheet_Polarizer_History}, paved the way to develop compact polarizers. The polarizer-based device industry has an 8.4$\%$ growth prediction through 2021-2027 indicating the necessity of high-caliber research to design wideband polarizers.  Hence,  an ultra-compact, low-loss broadband polarizer is always desired for transverse electric (TE) or transverse magnetic (TM) pass filter design and other diverse optical applications. Conventional polarizers are mainly classified into three types: absorption-based polarizer\,\cite{bludov2012tunable_graphene_polarizer}, reflection-based (Brewster-angle) polarizer\,\cite{zhupanov_2017_Brewster_angle_polarizer_ZrO2_SiO2} and refraction-based polarizer using prism\,\cite{Wu_applies_optics_2019_prism}. Polarizers can be used for versatile applications such as augmented reality, optical communication, optoelectronics, scientific instrumentation \cite{george2013_polarizer_application_scientific_instrumentation}, polarized light microscopy, and prediction of moisture contents in green 
peppers\,\cite{green_pepper_Faqeerzada_2020}. However, conventional polarizers are neither compact nor integrable to photonic circuits\,\cite{bao_2011_nature_broadband}. Moreover, most conventional polarizers suffer from lower damage thresholds and narrower operating temperature ranges \cite{zheng2021dichroic}. More importantly, they are expensive, incompatible with existing micro/nanofabrication technology, and unsuitable for monolithic integration\,\cite{yu2000_conventional_drawback}. Integration of nanoparticles offers distinctive characteristics in photonic and optoelectronic devices as their nanoscale dimensions provide optical and electronic confinement\,\cite{Nowshin23OptCon,Nowshin23RINP}. Nanoparticle-based polarizers are comprised of particles of a specific size, which filter out lights with a certain polarization orientation. Nanoparticle-based polarizers have advantages over polymer-based and conventional polarizers due to their tolerance to various external effects and in terms of flexibility\,\cite{Ferraro:16}. Metallic wire grids are the most utilized polarizer structure. These polarizers comprise many thin metal wires arranged parallel to each other. The wires act as a grating that reflects the polarized light along the direction of the wires while transmitting the polarized light perpendicular to the wires. Numerous studies have explored the adaptability and enhancement of such grids/wires in the NIR and THz range. Till now, most of the wire-grid polarizers suffer from operating range issues, and narrower bandwidth which limits the applicability of those polarizers\,\cite{zhao_2020_nanowire,shin_2013_Al_wiregrid,weber_2009_wiregrid_for_uv}.

\begin{figure*}[ht]
\centering
   \includegraphics[width=0.9\textwidth]{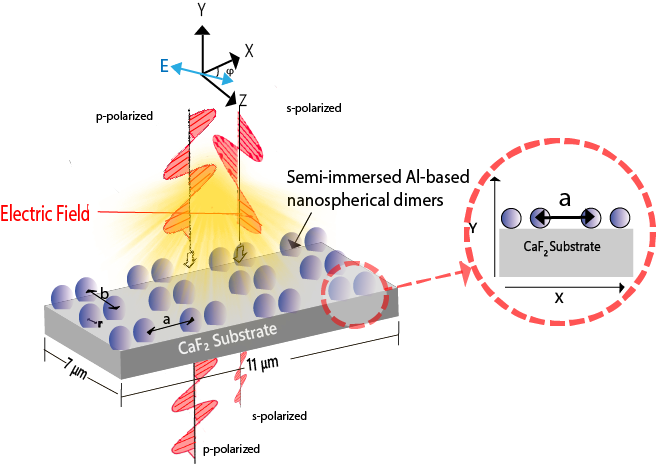}
    \caption{Proposed Al dimer nanosphere based polarizer. Incident light propagation is illustrated for p-polarization and s-polarization. The inset figure focuses on the semi-immersed Al dimer placement on CaF$_2$.}
    \label{fig:Device Structure}
\end{figure*}

To design a polarizer having light filtering capability on the NIR regime, it must have a nanometer feature size\,\cite{submicron_resonator_marzia_zaman2020}. Nano-imprint lithography and nano-transfer printing reveal new avenues of design exploration scopes for such polarizers. In visible and NIR regimes, cascaded grids comprised of Al have been proven effective for designing circular polarizers\,\cite{Fan_2023_Al_circular}. Particularly, the NIR range got much attention from researchers as the wavelengths used in telecommunication reside in the NIR range. Most of the proposed Al-based polarizers in the NIR regime had an extinction ratio of around 30 dB\,\cite{Hokari_2023_Al_triangular}. The THz regime is the least explored region of the electromagnetic spectrum. THz devices have diverse applications such as military applications, short-distance communication, non-invasive measurement, and space research. Now-a-days, THz range communication devices are becoming more important in military industries. Such military devices would benefit from compact THz polarizers. Therefore, there is a need for improvement of THz range polarizer performance. THz range polarizers are used to remove the unwanted components of the THz emitters to enhance performance sufficiently. Parallel metallic wires are spaced at sub-wavelength intervals upon a transparent substrate in THz wire-grid polarizers. This type of THz polarizer has versatile applicability, but the operating frequency span is limited for these wire-grid polarizers. Moreover, due to the fragility of such metallic wires, some groups worked on carbon nanotube-based fiber polarizer, which demonstrated up to 30 dB extinction ratio within a 0.5 dB maximum insertion loss\,\cite{a_zubair_apl_2016_cnt_polarizer}.  Zhou \textit{et al.} reported a near-ideal hydrogenated Ge-doped in-fiber polarizer by using a Bragg grating structure and showed a 99.99\% degree of polarization \cite{Zhou:05}. Middendorf \textit{et al.} reported the maximization of extinction ratio up to 78 dB by utilizing high fill fraction grids on substrate\,\cite{Middendorf_2014_high_ff_higj_er}. A facile solution process had been proposed by Kang \textit{et al.}  for fabricating Al wire grids for polarizer \,\cite{kang_aom_2018_Al_nanograting_02}. Such wire-grid polarizers have a sufficiently higher extinction ratio. However, these are dependent on various process parameters, such as temperature.  Li Q \textit{et al.} proposed a reflection-type augmented reality 3D display system that uses a reflective polarizer as both an imaging device and an image combiner. Employing metal particles instead of dichroic films allows desired absorption and scattering, hence, polarization in a broader range \cite{Metal_nanoparticle_zhang_2015}. In ideal polarizers, the extinction ratio is infinite, whereas it has a limited value in practical devices. Obtaining high transmission with a sufficiently high extinction ratio on a single device at the THz frequency regime is difficult. Hence, exploring newer techniques to develop near-ideal THz polarizers with higher extinction ratios and lower insertion loss is essential.

In the THz regime, several research works demonstrated promising results in on-chip polarization device development. Sarker \textit{et al.} proposed a tunable surface plasmon resonance-based terahertz (THz) polarizer with adjustable operating frequency\,\cite{dip_optica_polarizer}. On the other hand, in grain-like nanoparticle family, spheres, spheroids and dimers were investigated by numerous researchers. In nanoparticle-based polarizers, polarizing property was sensitive to the nanoparticle morphology\,\cite{ZHANG_JQSRT_2017_sphere_spheroid}. Intavanne \textit{et al.} reviewed recent progress of holographic applicability of polarizers using metasurface\,\cite{Intaravanne_2019_metasurface_review}.   Despite the above-mentioned approaches and recent development of on-chip TE/TM pass polarizers \cite{Bai_2019_on_chip_TE_pass}, there are scopes of further study of maximizing the elimination of unwanted polarization state from transmitted light in compact nanoparticle polarizers. 

\begin{figure*}[ht]
\centering
\includegraphics[width=1\textwidth]{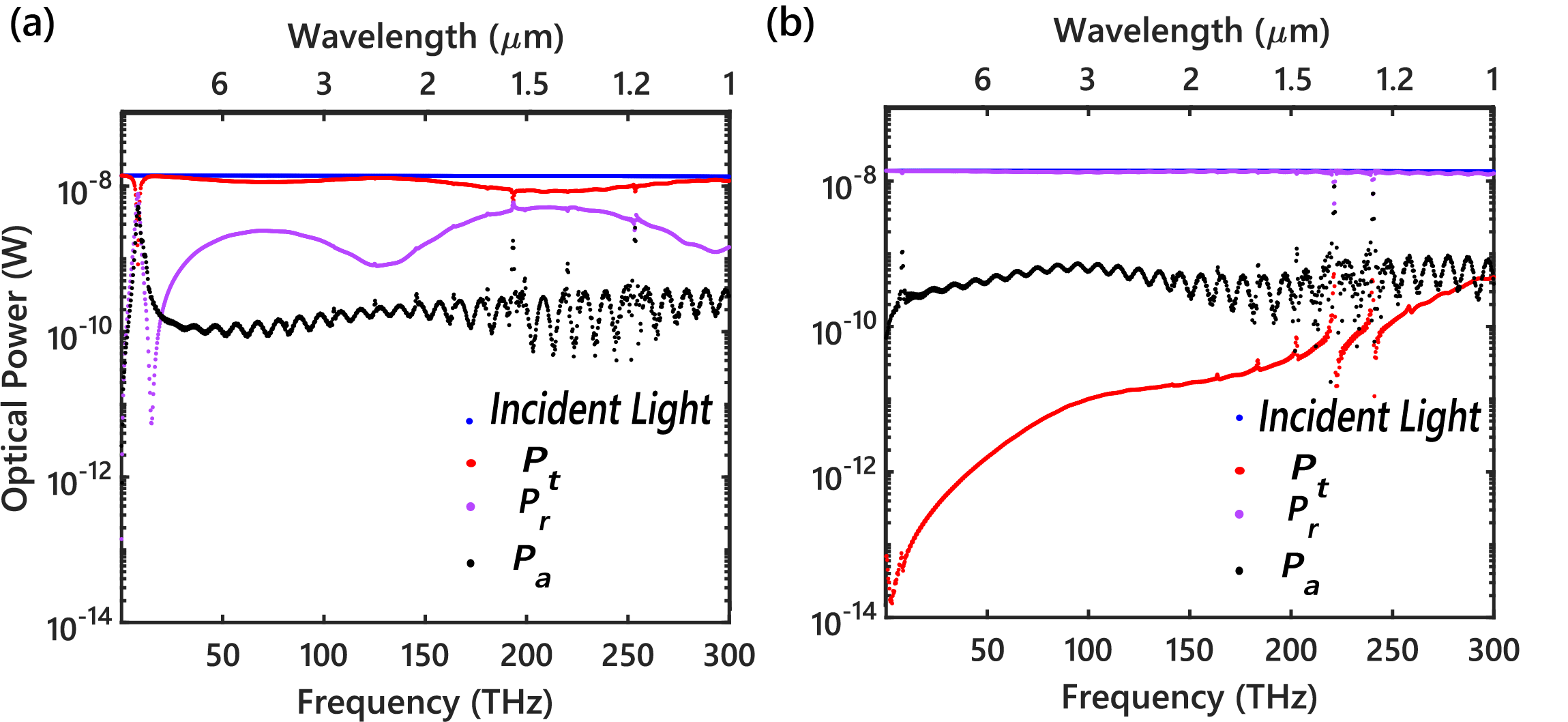}
\caption{Transmitted power ($P_t$), reflected power ($P_r$) and absorbed power ($P_a$) for (a) p-polarized incident light and (b) s-polarized incident light. The incident light source was tuned to emit a uniform optical power of $10^{-8}$ W for the studied wavelength regime.}
\label{Fig_RTA}
\end{figure*}

In this study, we proposed a polarizer that can effectively polarize light in both NIR and the comparatively less explored THz regime. Intensive optimization of nanoparticle morphology was conducted utilizing the  FDTD technique to obtain a broadband and high-extinction ratio. Alongside the extinction ratio, the insertion loss was also calculated which is one of the vital performance parameters for any polarizer regardless of the polarization mechanism, such as reflection, double refraction, and plasmonic polarization. We determined several performance parameters, such as the Mueller matrix index, transmittance-polar plot, and polarization ellipse, to evaluate our proposed polarizer and conducted a comparative study with previously reported polarizers. Findings from this work will be beneficial to fabricate an ultrabroadband polarizer.

\begin{figure*}[htb!]
\centering
    \includegraphics[width=1\textwidth]{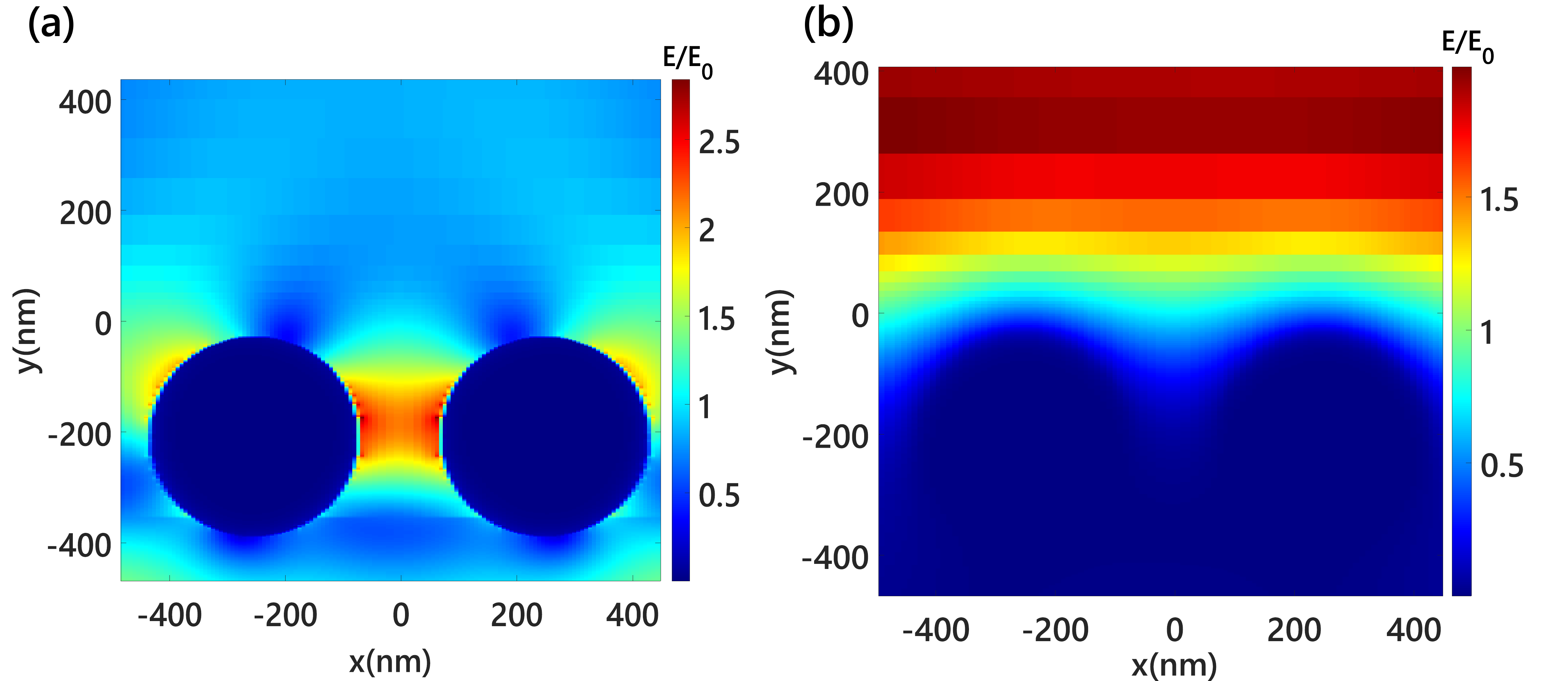}
    
\hfill
\caption{Color-map of electric field distribution at 3$\mu$m wavelength for (a) p-polarized and (b) s-polarized incident light. For s-polarized incidence, nearly zero transmission was observed through the dimer.}
\label{T_color_plot}
\end{figure*}

\section{Methodology}
In this study, we designed a polarizer comprised of a semi-immersed Al dimer nanosphere grid on a 15\textbf{ $\mu$}m CaF$_2$ substrate as can be seen in Fig.\,\ref{fig:Device Structure}. Ansys Lumerical FDTD was used to simulate the structure under illumination and calculate its optical properties. Using MATLAB, a numeric computing platform, we determined the performance-metric parameters utilizing the data extracted from the FDTD solver.  The proposed structure had dimer-shaped nanoparticles having a radius of 200 nm embedded into CaF$_2$ substrate. The proposed structure had dimer nanoparticles with a radius of 200 nm embedded into the CaF$_2$ substrate arranged in a 2D array. The lateral distance between two adjacent dimer spheres was 800 nm, denoted by a in Fig. \ref{fig:Device Structure}. Whereas the vertical distance has been noted as b.  Initially, our polarizer was comprised of uniformly arranged Al nanospheres. In that case, we got a high extinction ratio; however, the insertion loss was high. Hence, the trade-off between insertion loss and extinction ratio emerged, necessitating optimization and renovation of the nanoparticle arrangement for better transmission-blocking characteristics. Such arrangement resulted in a higher fill fraction on CaF$_{2}$ substrate. A higher fill fraction has some advantages in obtaining a high extinction ratio. The details of the structural optimization are discussed in {Supplementary Material}.

Material selection is a vital step for designing any nanophotonic device. We used metallic nanoparticles in our polarizer as they are free from the existing limitations of dichroic film polarizers, such as lack of tunability and operation range limitation \cite{Metal_nanoparticle_zhang_2015}. 
In the case of the substrate, we needed a transparent material in a wide wavelength regime. Commonly used transparent substrate materials include silica (SiO$_2$), magnesium fluoride (MgF$_2$), barium fluoride (BaF$_2$), zinc selenide (ZnSe). However, most of them are either hygroscopic or unstable \cite{gopalan2019_substrate_choose}. Organic polymers can be an alternative solution as many of these are transparent, cheap, and most importantly, flexible. However, they have several vibrational absorption peaks beyond the IR range, which is inconsistent with our device requirement. CaF$_{2}$ is a massively used substrate material having transparency from 0.15 $\mu$m to 9 $\mu$m and chemically stable. The CaF$_{2}$ crystal is optically isotropic, non-hygroscopic, and insoluble in most acids and alkali resistant \cite{TYDEX_CaF2}. These made it suitable for outdoor device applications such as windows and polarizers where degradation with time bears great importance. The only drawback of using  CaF$_{2}$ as substrate is its fragility, which is also applicable to other glass-based substrate families. \\

In the FDTD simulation, a plane wave light source was considered as the incident light source, and transmission and reflection monitors were placed in front of and behind the device. We considered the X-axis and Z-axis as the transmission and attenuation axes, respectively.  Transmittance and reflectance through the polarizer were measured using two separate monitors. A reflectance measuring monitor was placed behind the light source to prevent the inclusion of front-propagating light waves emitted from the light source. We calculated the emitted power from the incident plane wave source and transmitted power through dimers for both p and s-polarized light. In symmetric models such as our designed polarizer, symmetric/anti-symmetric conditions are efficient; however, they restrict data collection to only half of the simulation region, affecting the placement of monitors like time monitors. However, these conditions can complicate advanced analysis scripts such as polarization ellipse analysis. Therefore, to avoid misleading calculation results, we considered a perfectly matched layer (PML) on all the boundaries of the simulation arena. It increased Maxwell's equations-solving calculation's time and throughput requirement but resulted in accurate and reliable solutions. In PML, a generic material with high absorbance and little reflection was considered to ensure minimal reflection at the border edges of the structure. Despite the fact that a perfect layer should have zero reflectance, in this case, the abrupt change in material characteristics at the boundary led to a small amount of reflection at boundary edges.\\

We calculated the figure of merit parameters to evaluate the performance of our proposed polarizer and compare it with the commercial ones. Among them, Mueller matrix indices are vital. Mueller matrix is a $4\times4$ real-values matrix defining the polarization properties of a material. Among sixteen indices, the most significant three indices, m$_{12}$, m$_{22}$, and m$_{33}$ can be expressed as,

\begin{equation}
m_{12}=\frac{{T_{\parallel}}^{2}-{{T_{\bot}}^{2}}}{2}.
\label{m12}
\end{equation}

\begin{equation}
m_{22}=\frac{{T_{\parallel}}^{2}+{{T_{\bot}}^{2}}}{2}, 
\label{m22}
\end{equation}

\begin{equation}
m_{33}=T_{\parallel}T_{\bot}.
\label{m33}
\end{equation}

Here, T$_{\parallel}$ and T$_{\bot}$ represents the transmission through polarizer at parallel (x) and perpendicular (z) axis. Our findings utilizing these formulae are reported in later sections.\\

We conducted the optical simulation using completely s-polarized or p-polarized light so far. We placed a polarization-sensitive monitor to determine the intensity component of different polarization angles. However, we utilized a different approach to determine how much light got aligned with a specific polarization state. We used two separate light sources having polarization angles of 0$^\circ$ and 90$^\circ$, respectively. An s-p polarization filter was placed on the opposite side of the light source. This filter function determined the polarization ellipse for both linear and circular polarized light. It is always a good idea to check whether the filter works for applied boundary conditions. Therefore, we calculated the polarization ellipse in an empty simulation arena before investigating the proposed polarizer, which resulted in a straight line with a 45$^{\circ}$ slope and a circular curve for linear and circular polarized incident light. Afterward, we simulated the same setup with our polarizer to calculate the deviation from the ideal case.

\begin{figure*}[ht]
\centering
\includegraphics[width=1\textwidth]{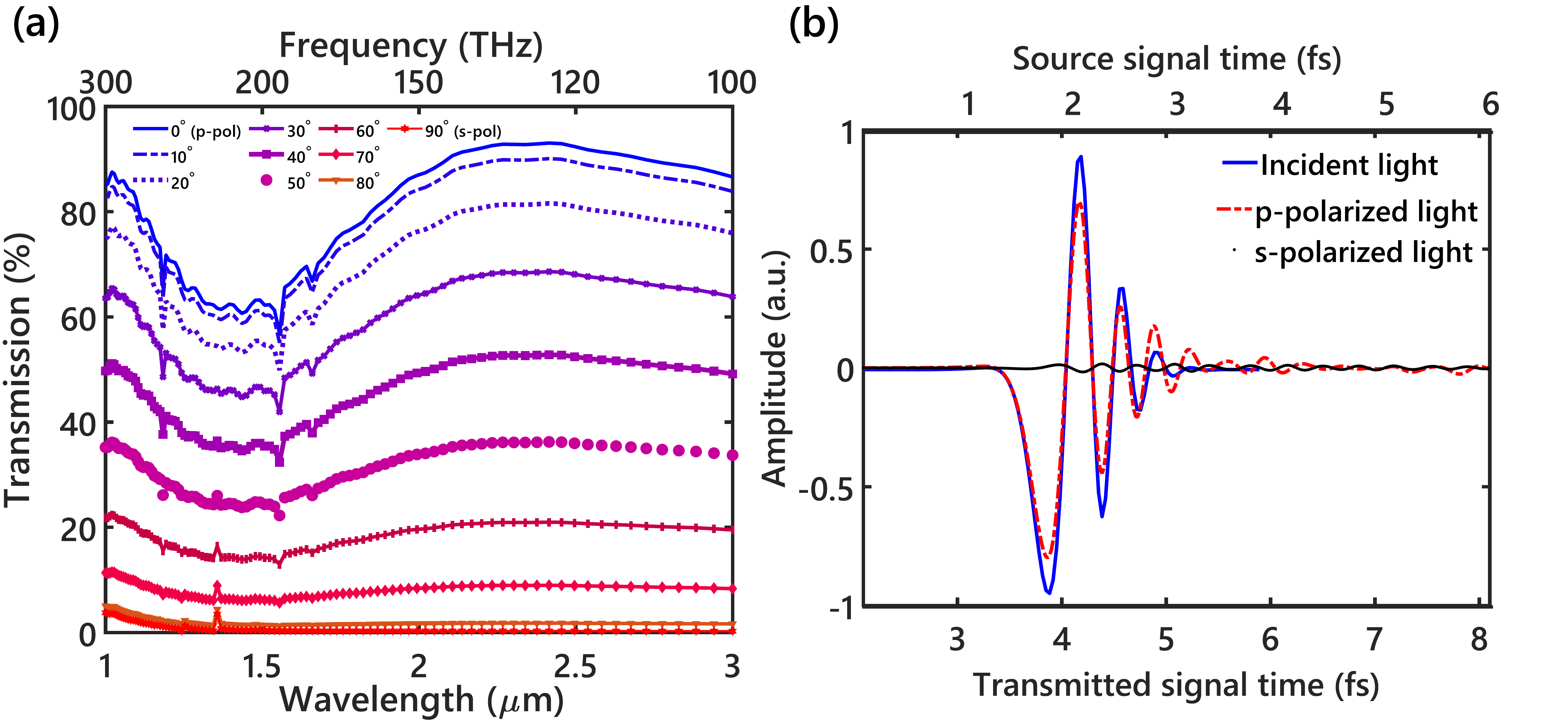}
\caption{(a) Gradual transmission blocking for increasing polarization angle. (b) Time domain signal of p and s-polarized light through the polarizer.}
\label{Tx_time_domain}
\end{figure*}

\section{Results and discussion}
\subsection{Ultra-broadband polarization-sensitive response}
We determined the transmission, reflection, and absorption optical power for the p and s polarized incident light. The resultant optical power spectra are shown in Fig. \ref{Fig_RTA}. The results shown in this figure and subsequent ones were calculated by simulating dimer-shaped nanoparticles arranged 1000 nm × 800 nm spaced array grid. The dimer spheres had a radius of 200 nm embedded into CaF2 transparent substrate.  In Fig. \ref{Fig_RTA}, red markers represent the transmitted power in watts, whereas the blue, black and violet ones represent the incident, absorbed and reflected power through the polarizer. We can see that the transmission line mostly overlaps with the incident power  indicates the maximum transmission for the p-polarized incident light. On the other hand, a sharp decrement of transmitted optical power in lower wavelength regime was found for s-polarized light. The transmitted light power attenuated to $10^{-11}$ from $10^{-8}$ W by just changing the incident light polarization angle. It implies that the proposed design will allow lights with a certain polarization angle to pass through when unpolarized light was incident on it. Reflection by Al nano-dimers was getting sufficiently larger when we changed the orientation of the incident light. This reveals the underlying mechanism of our proposed structure's polarization-dependent transmission modulation property. Apart from the transmission and reflection profile, our simulation results suggest that absorption profile of both p and s polarized light are in similar range ($\approx 10^{-10}$ W). This similar trend of absorption strongly supports the reflective mechanism of polarization in dimers.\\\textbf{
}

In nanoparticle-based polarizers, particles absorb or reflect incident radiation; if non-polarized light is applied on nanoparticles, the light will be polarized. In comparison, polarized light on the polarizer gets blocked or transmitted through the polarizer depending on its polarization orientation. To visualize the polarization characteristics of our proposed polarizer, we calculated the field profile for 3$\mu$m incident light. The findings from our calculation are shown in Fig. \ref{T_color_plot}. Almost all s-polarized incident light got reflected by Al dimers, whereas all were being transmitted for p-polarized incident light. Moreover, an electric field enhancement was observed at the inter-sphere spacing and sphere edges for p-polarized light. Here, the incident light was considered to be propagated from top to bottom of the structure shown in Fig.\,\ref{T_color_plot}. A video of electric field propagation for p-polarized and s-polarized light is provided in visualization 1 and visualization 2, respectively.  Transmission anisotropy through our dimer-based structure resulted in a high extinction ratio. A much better transmission-blocking property was observed for the proposed dimer-based polarizer than the sphere-based one. Minimal (or zero) spacing between dimers can enhance transmission-blocking properties, leading to promising photonic and optical device development\cite{Akhtary:23_dimer}.

\begin{figure*}[ht]
\centering
\hspace*{-0.8cm}
    \includegraphics[width=1\textwidth]{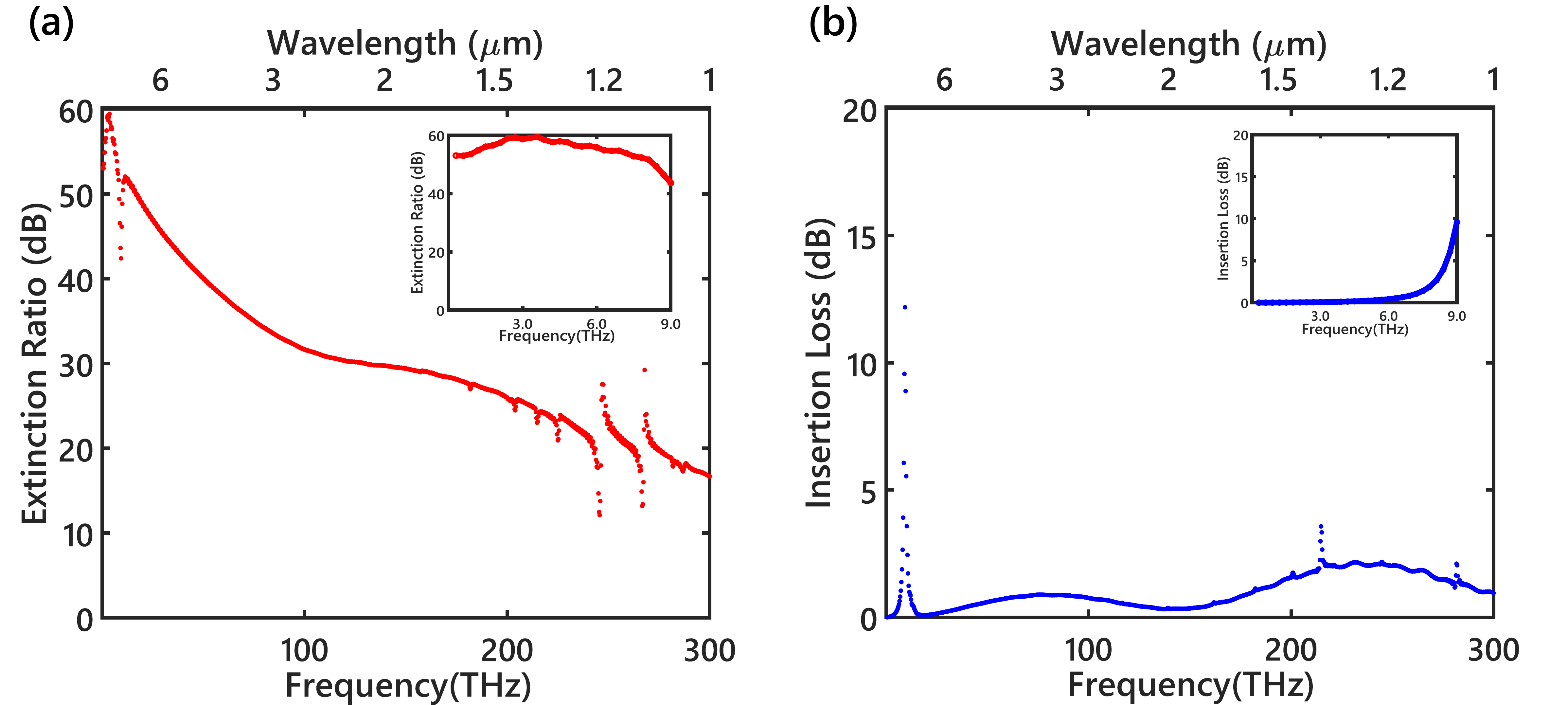}
\caption{(a) Extinction ratio, and (b) insertion loss  for wideband frequency range (0.18 THz to 300 THz).}
\label{THz_response_Spike_shift}
\end{figure*}

The proposed polarizer depicted a gradual transmission-blocking property with incident light polarization, as shown in Fig. \ref{Tx_time_domain} (a). However, a closer look at the insertion loss and extinction ratio characteristics can be found in Supplement 1. Fig.\ref{Tx_time_domain} (b) depicts the Al dimer polarizer's transmitted electric field in the time domain for different polarization angles. In Fig.\,\ref{Tx_time_domain}, the polarization sweep was conducted for a light source with 1 to 3 $\mu m$ wavelength range. However, our proposed polarizer followed similar gradual transmission-blocking properties for the whole spectrum studied, which is prominent from our calculated insertion loss. When the polarization angle ($\phi$) of the light was 90$^{\circ}$ (s-polarized light), the polarizer attenuated the light through mostly reflection mechanism. And, it is evident from the time domain waveforms of the transmitted electric field that s and p-polarized light were significantly modified (see Fig.\,\ref{Tx_time_domain} (b)). As the polarization angle of light increased, the transmitted electric field's amplitude significantly decreased.  Theoretically, in the case of our structure based on reflective nanoparticles\,\cite{zhao_2020_nanowire}, the s-polarized component of incident light is effectively reflected by the particles and, to a lesser extent, from the substrate. In contrast, since the p-polarized component is parallel to the dimer array, incident light propagated through the substrate but the high reflection from the nanosphere-based dimers strongly reduced the ultimate transmission. For a fixed substrate and nanoparticle material, the reflection of the s-polarized component is highly dependent on inter-sphere spacing. However, the dimers and the substrate have absorbance, which introduces lossy propagation. Thus, the overall transmission of our structure is slightly decreased by the non-negligible effects of absorption affecting both s- and p-polarized components of incident radiation.

We simulated the polarizer for 0.18 -- 300 THz (1 to 1665 $\mu$m) and determined the transmission, reflection, and absorption spectra. Our design criteria for the polarizer were to achieve the highest possible extinction ratio while keeping the insertion loss at a minimum level. Our analysis suggested 1000 nm horizontal spacing and 800 nm vertical spacing to obtain high transmission for the transmission axis and low transmission for the attenuation axis (see Supplement for details).

\subsection{Performance analysis}
There are various measures to determine a polarizer's figure of merits, such as extinction ratio, insertion loss, Mueller matrices, and polarization ellipse. In this study, we determined these parameters to evaluate the performance of our proposed structure. %Mechanical characteristics such as flexibility  crucial in determining polarizer performance. 
Gopalan \textit{et al.} reported a polarizer comprised of Au dipole arrays, where they showed that the optical characteristics were consistent after conducting more than 100 bending cycles\,\cite{gopalan2019_substrate_choose}. This experimental result suggested the possibility of using our Al dimer-based polarizers in flexible devices. Flexibility measurement and bending tolerance-related experiments of our proposed polarizer were beyond the scope of our theoretical study. 

\begin{figure}[ht]
\centering
\includegraphics[width=0.5\textwidth]{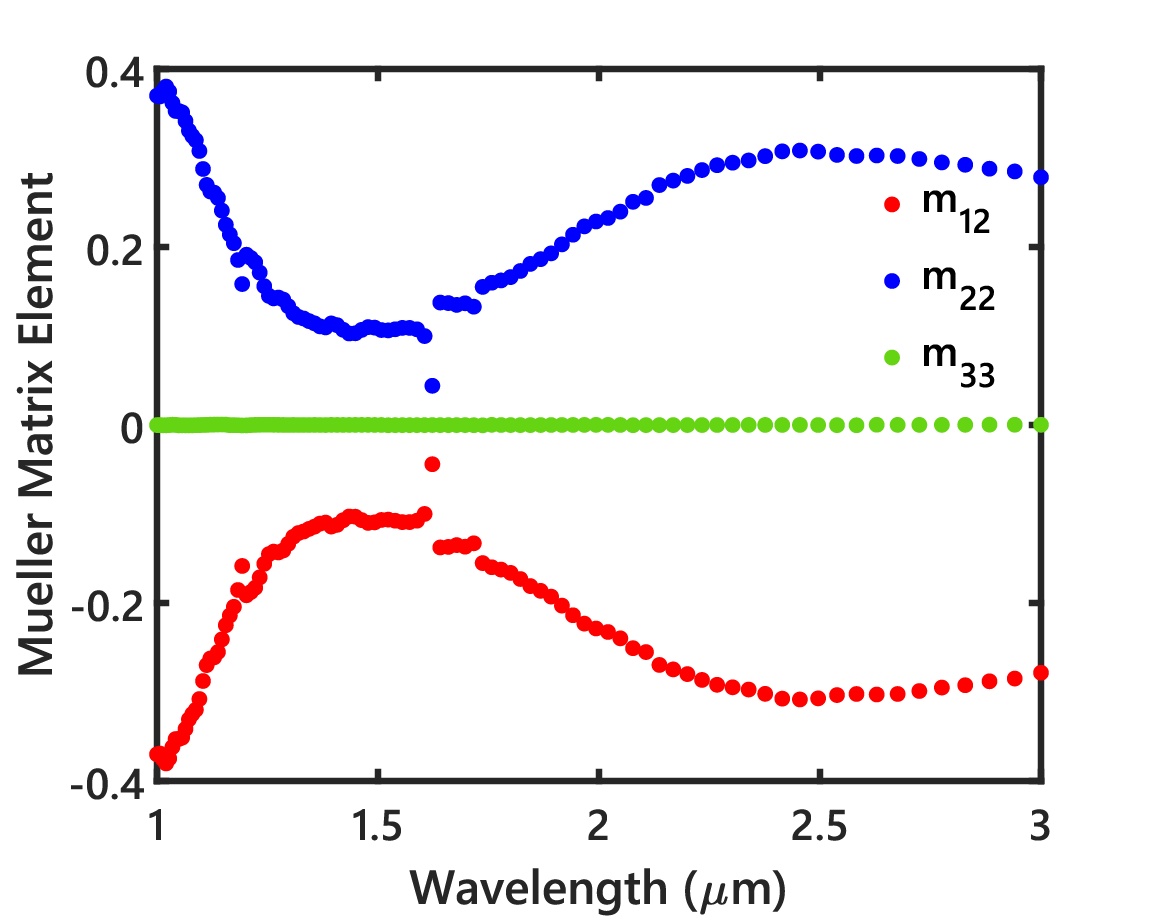}
\caption{Mueller matrix elements, $m_{12}, m_{22}, m_{33}$ determined for the proposed polarizer.}
\label{Mueller_matrix}
\end{figure}

\begin{figure*}[ht]
\centering
\includegraphics[width=1\textwidth]{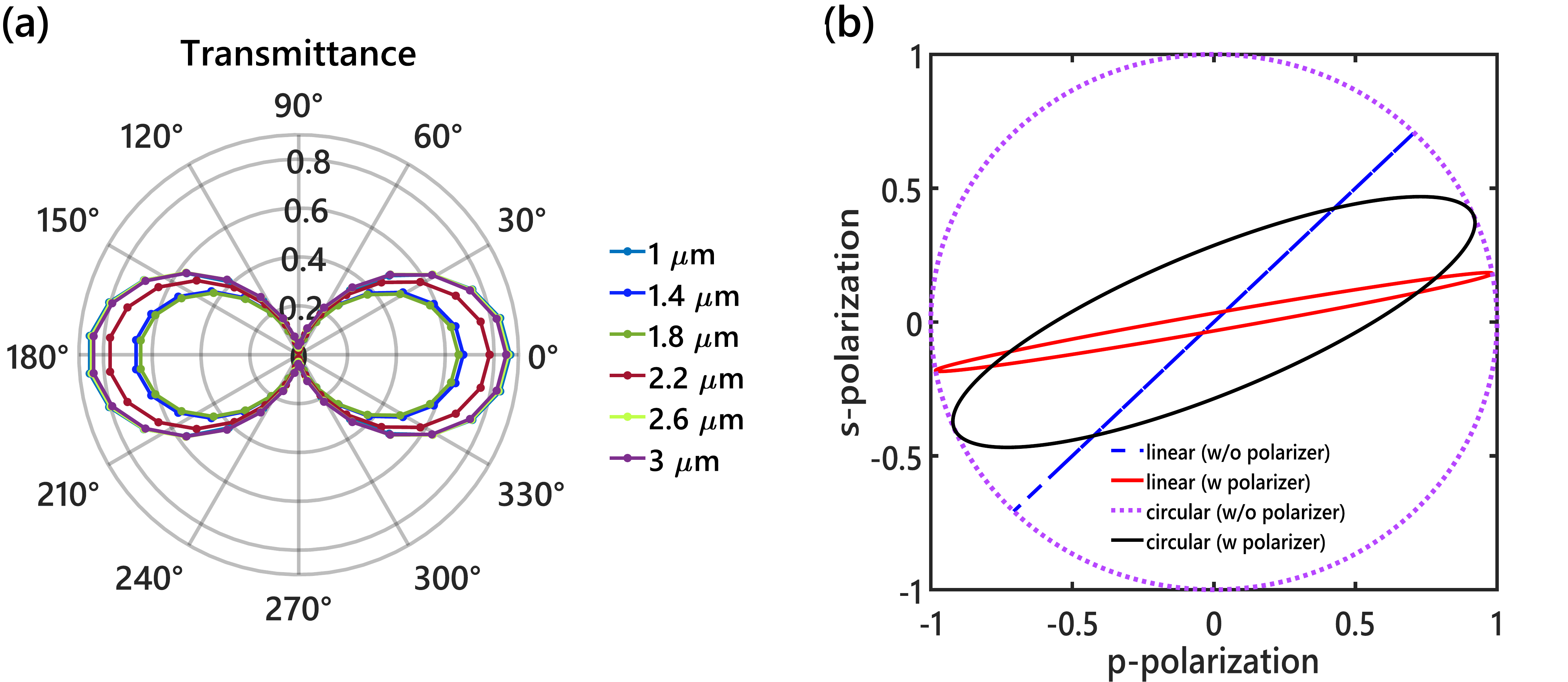}
\caption{(a) Polar plot of transmittance for NIR regime (1 -- 3 $\mu m$). (b) Polarization ellipse for linear and circular polarization without and with the proposed polarizer structure.}
\label{Polar_plot}
\end{figure*}

\subsubsection{Extinction ratio}
The extinction ratio, $ER$, which measures the degree to which light is confined in a principal linear polarization mode, is determined by\,\cite{a_zubair_apl_2016_cnt_polarizer},

\begin{equation}
\label{Extinction ratio}
ER(dB)=-10 log (\frac{P_{s}}{P_{p}}).
\end{equation}

Here, $P_{s}$ and $P_{p}$ are the optical power of attenuation and transmission axes, respectively. For linear polarizers, the transmission axis is the direction in which most light polarized in that direction is transmitted, and the attenuation axis refers to the light attenuation direction. In Fig. \ref{THz_response_Spike_shift} (a), the $ER$ is plotted for the wideband spectrum. Extremely high extinction ratio of more than 10$^5$ can be achieved by our proposed polarizers. The average extinction ratio for the THz regime was found to be around 50 dB. On the other hand, it was found as 23.33 dB in the NIR regime,  from 1 to 3 $\mu m$  (see Supplementary document). However, for whole wavelength regime studied (1 to 1665 $\mu$m) was 29.01 dB .  To maximize the extinction ratio, intensive optimization of the geometry parameters of the dimer was conducted, including space reduction between dimer, and sphere radius decrement. However, the insertion loss got higher in that case, which worsened the Mueller matrix values. Moreover, compared with non-metallic polarizers, metallic nanoparticle-based polarizers such as our proposed one had higher insertion loss. Therefore, a compromise to meet both demands was the most viable solution. More detail of this optimization study is described in Supplemental document. A high extinction ratio demanded a large fill fraction of nanoparticles on the substrate, resulting in a higher loss. Therefore, an optimal extinction ratio with minimal insertion loss was designed for metallic reflection-based polarizers.

\subsubsection{Insertion loss}
Insertion loss, $IL$ is the loss of optical power while it passes through a polarizer. When a light signal passes through a polarizer, it is modulated according to the polarizer structure or material characteristics. Due to absorption and other lossy events, the incident light gets attenuated, which varies with frequency and polarization angle. Mathematically, insertion loss is expressed in dB as\,\cite{a_zubair_apl_2016_cnt_polarizer},

\begin{equation}
\label{Insertion Loss}
IL(dB)=-10 log (\frac{P_{t}}{P_{i}}).
\end{equation}

Here, $P_{t}$ is the power of light transmitted through the device, and $P_{i}$ is the power of incident light. To evaluate the performance of the proposed polarizer, we determined the insertion loss in dB shown in Fig.\,\ref{THz_response_Spike_shift} (b). A maximum of 2.3 dB insertion was observed in the NIR region, which was around 40\% incident power (for a more detailed figure, see Supplemental). It is higher than other non-metallic polarizers; however, it is comparable to polarizers with metallic nanoparticles\,\cite{Metal_nanoparticle_zhang_2015, abd_2022_ultra_compact_bi_metalic}. Our proposed nanoparticle polarizer has an average insertion loss of 1 dB over 1 to 1665 $\mu$m wavelength range. This loss of incident light for p-polarization can be understood from the time domain perspective as shown in Fig.\,\ref{Tx_time_domain} (b), where the p-polarized light faced low attenuation, and that is why the polarizer was not fully transparent in p-polarized light. We optimized the horizontal and vertical spacing between the dimers and determined the optimal values to extract the maximum extinction ratio within minimal insertion loss. We simulated dimer-shaped nanoparticles arranged in an 1000 nm $\times$800 nm spaced array grid having a radius of 200 nm embedded into CaF$_2$ transparent substrate. Along with dimer spacing, we also investigated the effect of dimer radius on insertion loss and extinction ratio. Our observations on these optimizations are described in the Supplementary Material.

\subsubsection{Mueller matrix elements, polar plot, and polarization ellipse}
Mueller matrix is a standard form of representing the polarization states. This measure is more critical for polarizers with more than one polarizing element. Mueller matrices are 4$\times$4 matrices. Here, the polarization of transmitted light is expressed by a linear combination of four parameters, known as Mueller indices, as given by.
\begin{equation}
\begin{pmatrix}
S_{0}^{'} \\
S_{1}^{'} \\
S_{2}^{'} \\
S_{3}^{'}
\end{pmatrix}= \begin{pmatrix}
m_{00} & m_{01} & m_{02} & m_{03} \\
m_{10} & m_{11} & m_{12} & m_{13} \\
m_{20} & m_{21} & m_{22} & m_{23} \\
m_{30} & m_{31} & m_{32} & m_{33}
\end{pmatrix}\times
\begin{pmatrix}
S_{0}\\
S_{1}\\
S_{2}\\
S_{3}
\end{pmatrix}.
\label{Mueller}
\end{equation}

In this study, we calculated $m_{12}$, $m_{22}$, and $m_{33}$ 
indices of the Mueller matrix plotted in Fig. \,\ref{Mueller_matrix}. Here, $m_{33}$ followed the ideal value of the index, 0. However, $m_{12}$ and $m_{22}$ did not fully align with the ideal values (+0.5 and -0.5, respectively) near 1.5 $\mu m$ wavelength region. The sudden fluctuation of reflectance and spike in insertion loss spectra at that wavelength was responsible for such deviation.\\

Fig.\,\ref{Polar_plot} (a) shows the polar plot of the normalized transmitted power, T as a function of polarization angle,  $\theta$. The values of T were obtained in the same manner as described in the methodology section. The difference between minimum and maximum transmission was found to be more than 80$\%$, which is relatable with transmission spectra. We recorded polar diagram data for over ten wavelength points between 1 and 3 $\mu m$. Selected plots are presented in Fig.\ref{Polar_plot} (a). It can be noted that all the frequency points did not support 80$\%$ transmission deviation. A plausible reason for this difference might be the absorption introduced by dimer metals, resulting in non-ideal elliptical polarization described below.

Polarizer elements have the property of altering the transmittance of light through it based on polarization orientation. The proposed polarizer consisted of a semi-immersed Al dimer grid. The previously discussed findings showed that the nanoparticle grid reflected s-polarized light and transmitted the p-polarized one through it. A polarization ellipse is a measure of the polarizability of a polarizer. Moreover, it also explains the circular polarizability of any polarizer. When a monochromatic light is applied on the polarizer, it splits into two orthogonal eigenwaves. To incorporate this scenario, we used two plain wave sources with different polarization states to calculate the polarization ellipse for circular polarization\,\cite{PMMA_Birefringence_Tarjanyi_2019}. This initiated the production of circularly polarized radiation where the superposition of two plane waves occurred. They had the same amplitude, same frequency, and co-related phase angles. Before imposing the light into the proposed structure, we applied the source to free space to confirm the accuracy of the placed monitor. This test run gave us straight line for linear polarized sources and a circular response for circularly polarized light (shown as dashed curves in Fig. \,\ref{Polar_plot} (b). Afterward, we applied the light sources on our proposed polarizer and got sufficiently closer results to the ideal one (shown as solid curves in Fig. \,\ref{Polar_plot} (b). The deviation between the dashed and solid curves in Fig. \,\ref{Polar_plot} (b) can be attributed to the internal absorption of reflected lights throughout the dimer structure.

\begin{figure*}[ht]
\centering
\includegraphics[width=0.9\textwidth]{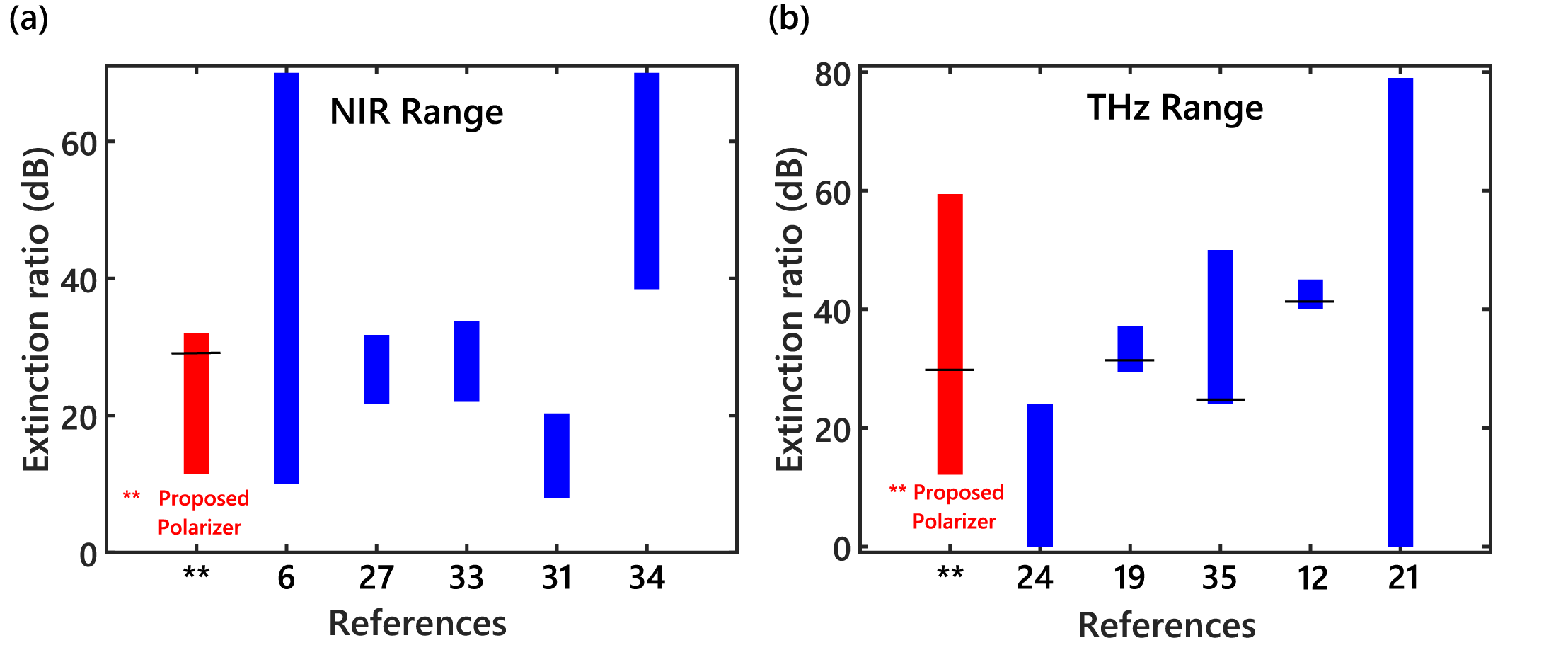}
\caption{Comparative candle plot of extinction ratio (in dB) extracted from extensive literature review, (a) for NIR regime and (b) for THz regime 
\cite{green_pepper_Faqeerzada_2020}, 
\cite{Bai_2019_on_chip_TE_pass},
\cite{jeon_2022_bilayer},
\cite{abd_2022_ultra_compact_bi_metalic}, 
\cite{inc_2020_commercial_polarizer},
\cite{dip_optica_polarizer},
\cite{a_zubair_apl_2016_cnt_polarizer}, 
\cite{Yamada:09},
\cite{Ferraro:16},
\cite{Middendorf_2014_high_ff_higj_er}.} 

%\cite{ZHANG_JQSRT_2017_sphere_spheroid}, 
%\cite{Akhtary:23_dimer}, 
%\cite{kim2015_nano_metalgrid}, .}
\label{candle_plot}
\end{figure*}

%\section{Discussion}
%%\label{}
\section{Comparative analysis}

Wire grid polarizers are commonly used for infrared applications, as they have several advantages over other types of polarizers, such as thin and lightweight design, easy integration with other optical components, wide operating wavelength range, larger acceptance angle, high thermal stability, higher laser damage threshold, low absorption, and high transmission\,\cite{zhao_2020_nanowire, shin_2013_Al_wiregrid, weber_2009_wiregrid_for_uv}. However, these polarizers have major limitations such as fragility, limited ER, and narrow operating bandwidth. Moreover, a narrower grid pitch is required to increase the ER for wire-grid polarizers, which necessitates higher precision in fabrication. To overcome these shortcomings, alternative structures such as nanosphere, ellipsoid nanoparticles, and bilayer arrangement have been proposed by several research groups\,\cite{jeon_2022_bilayer}. Our proposed polarizer can be compared to metallic wire-grid and nanoparticle polarizers based on several performance parameters. In Table \ref{Comparative_analysis}, we reviewed the characteristics and figure of merit of our proposed polarizer with the reported ones. The nanoparticles were uniformly distributed for all the nanoparticle-based polarizers mentioned in Table\,\ref{Comparative_analysis}.  Bi-metallic TM-passed polarizer reported by Elkader \textit{et al.} had a lower insertion loss. Utilizing a bi-metallic configuration comprised of aluminum zinc oxide (AZO) and ZrN, they obtained a 20.3 dB extinction ratio with 0.14 dB insertion loss\,\cite{abd_2022_ultra_compact_bi_metalic}. However, their device structure was complex which resulted in possible difficulties in the fabrication process. We optimized our dimer structure to have an extinction ratio as high as 50 dB in the THz range which was possible by proper metal selection and utilizing semi-immersed metal nanoparticles on a transparent substrate. Bao \textit{et al.} proposed another TE pass polarizer for the visible and near-infrared region, where the extinction ratio was found between 24 to 33.7 dB\,\cite{bao_2011_nature_broadband}. The promising attribute of their proposal was the compatibility of their structure with silicon-on-insulator fabrication technology. Metal selection is one of the vital steps in reflection-based polarizers. Ag has been treated as an attractive metal for designing polarizers. Moiseev \textit{et al.} demonstrated a reflection-based polarizer structure comprised of Ag spheres\cite{moiseev_2011_ag_sphere}. %They focused on the comparative study of different solvers by calculating reflectance, and transmittance in visible wavelength regime (560nm-680nm). 
Up to 80\% transmission with viewing angle consistency was achieved by utilizing the angled-evaporation method on short-period Al wire grids\,\cite{shin_2013_Al_wiregrid}. Weber \textit{et al.} presented an excellent comparative analysis of several material performances on wire-grid or similar reflection patterns. Their report was based on material performance comparison and extinction ratio in dependence of the grating height and duty cycle which suggested that aluminum outperformed in the whole spectral range which motivated us to select Al as the reflecting metal for our polarizer \cite{weber_2009_wiregrid_for_uv}. A similar Al-based polarizer was proposed by Kim \textit{et al.}, where they claimed a maximum extinction ratio of 10$^6$ at 600 nm wavelength. Their study focused on the deposition depth of Al nanoparticles on poly-ethylene phthalate (PET) substrate and got maximum polarizing performance at 70 nm depth of metal grid \cite{kim2015_nano_metalgrid}. Their findings suggest a great dependency of polarizer performance parameters on deposition depth. In our design, we considered such dependency and designed our Al dimers in a semi-immersed fashion, which helped us to obtain maximum polarization anisotropy on transmission and reflection properties. In the THz regime, numerous comprehensive studies motivated us to develop our polarizer a wideband-compatible one. Among them, Yamada's terahertz wire-grid polarizer consisting of micrometer-pitch Al grating on Si substrate by photolithography and wet etching \,\cite{Yamada:09}, Ferraro's Al grating-based polarizer on the sub-wavelength thin flexible and conformal foil of the cyclo-olefin Zeonor© polymer\,\cite{Ferraro:16} are prominent ones. Our findings after reviewing various reports of NIR and THz range polarizers have been summarized in candle plots of Figs\,\ref{candle_plot} (a) and\,\ref{candle_plot} (b) respectively. Our proposed polarizer (red bar) had an average extinction ratio of 23.37 dB for NIR regime (Fig.\ref{candle_plot} (a)). On the other hand, it was found to be more than 55 dB for the THz regime (Fig.\ref{candle_plot} (b)). The comparative bar chart suggests the great acceptability of our proposed one in broadband applications, especially in the THz regime.

\section{Proposed fabrication method} 

{Our proposed nanoparticle polarizer can be fabricated by embedding prolate ellipsoids on transparent substrate such as CaF$_2$. Farmer \textit{et al.} proposed a similar metal-nanoparticle array fabrication scheme utilizing Au spheres on nano-dimpled Al substrate\,\cite{farmer_2022_fabrication}. Similar approach can be considered in our proposed structure fabrication where Al spheres on CaF$_2$ substrate dimples can be placed to form semi-immersed dimers. Initially, the substrate can be etched in a patterned fashion to make dimples to place Al nanodimers. Array of nanodimers can be fabricated on a seperate membrane and then placed to CaF$_{2}$ substrate using synthetic resins such as polymethyl 2-methylcrylate (PMMA). During this transfer procedure, PMMA can be spin-coated on Al dimers. This step ensured that the Al nanodimers be uniformly covered with PMMA layer. The solidified portion of the PMMA-coated nanoparticle array can be detached carefully  with precise cutting instrument. It will be convenient to perform the whole process immersed in deionized (DI) water. Immersion allowed the DI water to seep in between the PMMA layer and the indented surface, enabling the PMMA layer to detach smoothly from the growth base. Afterwards, the dimers can delicately be lifted from the DI water and placed on the chosen substrate, which brought the nanoparticles within the PMMA layer into direct contact with the substrate. PMMA layer can easily be dissolved by placing a few drops of dichloromethane onto the substrate. Any remaining residue on the substrate was cleaned off with a rinse of acetone and DI water. Fig.\,\ref{fabrication_steps} illustrates the proposed fabrication steps.

\begin{figure}[ht]
\centering
\includegraphics[width=0.45\textwidth]{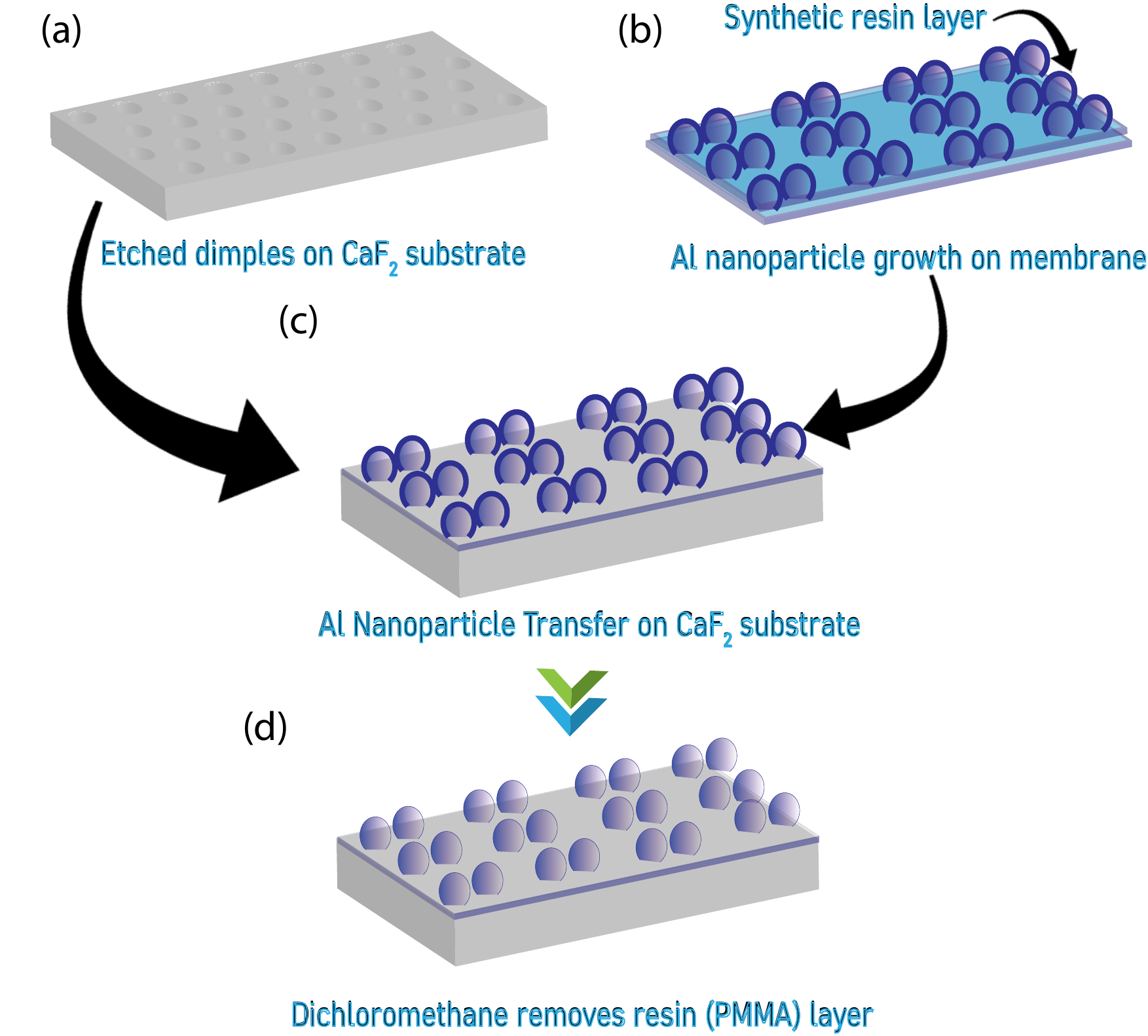}
\caption{Proposed fabrication steps-- (a) initial Caf$_2$ substrate etching, (b) Al nanoparticle formation on seperate membrane, (c) Al nanoparticle transfer to substrate using synthetic resin layer (d) dichloromethane removes the thin layer of resin such as PMMA from nanoparticle surface, lefts the solid Al nanodimers only.}
\label{fabrication_steps}
\end{figure}

\begin{table*}
    \centering
\caption{Comparative analysis of NIR and THz range polarizers}

\label{Comparative_analysis}
    \begin{tabular}{ccccc}
    \hline 
         Structure and dimension&  Insertion loss (dB)&  Extinction ratio (dB)&  Operating range&  Ref\\ \hline 
\makecell{Silicon hybrid\\plasmonic grating}& $<$4.5& --& \makecell{Visible and NIR\\(0.38$\mu$m--0.44$\mu$m and 1.52--1.58 $\mu$m)}& \cite{Bai_2019_on_chip_TE_pass}\\ 
 Bi-metallic compact& $<$0.9& --& NIR (1.94--2.06$\mu m$)& \cite{abd_2022_ultra_compact_bi_metalic}\\ 
         Bi-layer Au grid&  $<$0.17&  $<$25.5&  NIR (3--5$\mu m$)&  \cite{jeon_2022_bilayer}\\ 
         \makecell{Commercial polarizer\\from Thorlabs}&  $\sim$0.66&  41.25&  NIR (1--3$\mu$m) &  \cite{inc_2020_commercial_polarizer}\\
         \makecell{Al grating on flexible\\polymer foil}&  $<$1&  40-45&  THz (0.5--2.5 THz)&  \cite{Ferraro:16}\\ 
         Metallic wire-grid&  --&  50.83&  THz ( 0.2--1 THz)&  \cite{Middendorf_2014_high_ff_higj_er}\\
         \makecell{Surface plasmon in\\ graphene nanoribbon}&  1.87&  30&  THz (1.82--3.75 THz)&  \cite{dip_optica_polarizer}\\ 
         Carbon nanotube fiber&  $<$0.5&  30&  THz (0.2--1 THz)&  \cite{a_zubair_apl_2016_cnt_polarizer}\\ 
         Al dimer&  1.15&  23.37&  NIR (1--3 $\mu m$)&  This work\\
         Al dimer&  1.023&  55.03&  THz (0.18--9 THz)&  This work\\ \hline
    \end{tabular}

\end{table*}

\section{Conclusion}
We investigated the scopes and possibilities of developing a wide-band compact polarizer by employing Al-dimers. Our main goal was to propose a newer scheme for developing broadband metal-based polarizers that can effectively work from NIR to THz regions. In this study, the CaF$_2$ slab containing uniformly spaced Al semi-immersed dimers was successfully modeled, and the polarizer performance indicator parameters such as extinction ratio, insertion loss, Mueller matrix elements, and polarization ellipse were reported here. Our proposed polarizer showed a $>$55 dB extinction ratio in the THz regime with an average insertion loss of $\sim$1 dB. Mueller matrix element m$_{33}$ was found to be 0 for an ultrabroadband wavelength regime. The polarization ellipse of the proposed polarizer suggested its great usability in both linear and circular polarized light. Compared to the conventional ones, the impressive features of the proposed metal nanoparticle-based polarizer will extend the scope of research and development of thin-film optical filtration devices and photographic devices. Not limited to military devices for communication and sensing, a wide variety of polarization-based devices can be developed utilizing our polarizer with a high extinction ratio over a broad frequency range.

\section*{Acknowledgements}
M.S.I. and A.Z. acknowledge the support and facilities
received from the Department of EEE, Bangladesh University of Engineering and Technology (BUET).

\section*{Data Availability}
All relevant data that support the findings of this study are presented in the manuscript
and supplementary material file. Source data are available from the corresponding
author upon reasonable request.

\section*{CRediT authorship contribution statement}
\textbf{Md. Shariful Islam}: Conceptualization, Methodology, Visualization, Software, Investigation, Writing - Original Draft.
\textbf{Ahmed Zubair}: Conceptualization, Methodology, Visualization, Resource, Supervision, Writing - Original Draft, Writing-Review \& Editing.

%\balance

\bibliographystyle{ieeetr}
\bibliography{main}
\end{document}